\documentclass[superscriptaddress,nofootinbib]{revtex4}

\usepackage{graphicx}
\usepackage{axodraw}
\newcommand{\tplus}{T^+}
\newcommand{\tminus}{T^-}
\newcommand{\tpm}{T^{\pm}}
\newcommand{\tzero}{T^0}
\newcommand{\tzerobar}{\bar{T}^0}

\begin{document}
% You should use BibTeX and revtex.bst for references
\bibliographystyle{revtex}

% number  and your conference paper identification number on the
% title page in preprint mode. Multiple \preprint commands are allowed.
%\preprint{SNOWMASS 01/XXX}
%\preprint{SCIPP 01/33}

\title{Report of the Subgroup on Alternative Models and New Ideas}
% Optional argument for running titles on pages
%\title[]{}



\author{M. Chertok$^*$}
%\email[]{chertok@ucdavis.edu}
%\thanks{}
\affiliation{University of California at Davis, Davis, CA 95616}

\author{K. Dienes}
\affiliation{University of Arizona, Tucson, AZ 85721}

\author{S. Godfrey$^*$}
\affiliation{Ottawa-Carleton Institute for Physics,
Department of Physics, Carleton University, Ottawa, Canada K1S 5B6}

\author{P. Kalyniak}
\affiliation{Ottawa-Carleton Institute for Physics,
Department of Physics, Carleton University, Ottawa, Canada K1S 5B6}

\author{D. Kaplan$^*$}
\affiliation{University of Chicago, Chicago, IL 60637}

\author{G.D. Kribs}
%\email[]{kribs@pheno.physics.wisc.edu}
%\thanks{}
\affiliation{Department of Physics, University of Wisconsin,
             Madison, WI 53706}

\author{R. Mohapatra}
\affiliation{University of Maryland, College Park, MD 20742}

\author{Y. Nomura}
%\email[]{yasunori@thsrv.lbl.gov}
%\thanks{}
\affiliation{Department of Physics, University of California, Berkeley,
CA 94720}

\author{W. Orejudos}
%\email[]{orejudos@fnal.gov}
%\thanks{}
\affiliation{Lawrence Berkeley National Laboratory, Berkeley, CA 94720}

\author{N. Romanenko}
\affiliation{Ottawa-Carleton Institute for Physics,
Department of Physics, Carleton University, Ottawa, Canada K1S 5B6}

\author{M. Schmaltz}
\affiliation{Fermi National Accelerator Laboratory, Batavia, IL 60510}

\author{B. Schumm$^*$}
%\email[]{schumm@scipp.ucsc.edu}
%\homepage[]{http://scipp.ucsc.edu/~schumm}
%\thanks{}
\affiliation{Santa Cruz Institute for Particle Physics and the University of
   California at Santa Cruz, Santa Cruz, CA 95064}

\author{S. Su}
%\email[]{shufang@theory.caltech.edu}
%\thanks{}
\affiliation{California Institute of Technology, Pasadena, CA 91125}


\date{\today}

\begin{abstract}
We summarize some of the work done by the P3 subgroup on Alternative 
Models and New Ideas.  The working group covered a broad range of 
topics including a constrained Standard Model from an extra
dimension, a discussion of recent ideas addressing the strong CP problem,
searches for doubly charged higgs bosons in $e\gamma$ collisions, and 
an update on discovery limits for extra neutral gauge bosons at hadron 
colliders.  The breadth of topics reflects the many ideas and 
approaches to physics beyond the Standard Model.
\end{abstract}
% \pacs{}

%\maketitle must follow title, authors, abstract and \pacs
\maketitle

\noindent
$^*$ Subgroup conveners


\section{Introduction}

The search for physics beyond the Standard Model has become the
``Holy Grail'' of particle physics for the last twenty years.  In
addition to supersymmetry and dynamical symmetry breaking, which were
examined by other groups at this workshop, there is a plethora of 
ideas to fulfill this quest.  The most exciting new idea in recent 
years is the postulate of extra dimensions in addition to the standard 
four spacetime dimensions we are used to.  In this context,
Chertok {\it et al} 
examined the implications of a constrained Standard Model originating
in five dimensions.  In this model, the entire SM is supersymmetrized 
in five dimensions which leads to many new particles including a 
mirror set of supermultiplet fields and KK towers.  A compelling 
feature of this model is that it makes an unambiguous prediction for 
the physical Higgs boson of $m_h=(127\pm8)$~GeV. In the next section 
we describe some of the phenomenology and experimental prospects for 
this model.  Mohapatra and coworkers also took advantage of the possibility 
of  extra dimensions to provide new approaches towards solving the 
strong CP problem, a longstanding problem of the Standard Model. This,
along with some other ideas about the strong CP problem are summarized in 
section III of this report.  In addition to these new developments in 
particle theory, more ``conventional'' variants of physics beyond the 
Standard Model were studied at the workshop.  One of the most
straightforward extensions of the Standard Model is the existence of
larger Higgs representations.  These arise naturally in the Left-Right 
symmetric model, for example, in addition to other models.  A study 
using the $e\gamma$ mode of a high energy $e^+e^-$ collider to search 
for doubly charged Higgs is summarized in
section IV.  Another
common feature of many extensions of the Standard Model is the
existence of extra neutral gauge bosons.  Their discovery would have 
important implications  for what lies beyond the SM. An update to 
discovery limits for the hadron colliders discussed at Snowmass is 
given in section V.

\section{A Constrained Standard Model from an Extra Dimension}

This section represents a synopsis of a submission, Ref.~\cite{susy5d},
appearing in these proceedings.

Recently, a new approach to low energy supersymmetry breaking has been
given by Barbieri, Hall, and Nomura~\cite{BHN}.  Unlike the usual
approach of postulating a low energy effective theory with so-called
``soft'' supersymmetry breaking terms added by hand, the entire SM
is supersymmetrized in \emph{five} dimensions.
This means there are not only complete (${\cal N}=1$ in 4D) supermultiplets
consisting of the SM fields and their superpartners, but also a
``mirror'' set SM fields
and their superpartners.  This is required since supersymmetry
in 5D has double the number of supercharges than in 4D; i.e.,
an ${\cal N}=1$ supermultiplet in 5D has the field content of
a single ${\cal N}=2$ supermultiplet in 4D, which is equivalent
to \emph{two} ${\cal N}=1$ supermultiplets in 4D.

\begin{figure}
\begin{center}
\begin{picture}(350,90)(-10,-10)

  \Line(5,0)(320,0)
  \LongArrow(10,-10)(10,80)
%  \Text(0,85)[b]{mass}
  \Text(-40,30)[b]{mass}
  \Text(0,0)[r]{$0$}
  \Line(8,20)(12,20)    \Text(6,20)[r]{$1/R$}
  \Line(8,40)(12,40)    \Text(6,40)[r]{$2/R$}
  \Line(8,60)(12,60)    \Text(6,60)[r]{$3/R$}
%  \Text(60,90)[b]{$\psi_{M}, \phi_{H}, A^{\mu}$}
  \Text(60,80)[b]{$\psi_{M}, \phi_{H}, A^{\mu}$}
  \Line(40,0)(80,0)      \Vertex(60,0){3}

  \Line(40,40)(80,40)    \Vertex(60,40){3}
%  \Text(130,90)[b]{$\phi_{M}, \psi_{H}, \lambda$}
  \Text(130,80)[b]{$\phi_{M}, \psi_{H}, \lambda$}

  \Line(110,20)(150,20)    \Vertex(130,20){3}
  \Line(110,60)(150,60)    \Vertex(130,60){3}
%  \Text(200,90)[b]{$\phi^{c}_{M}, \psi^{c}_{H}, \psi_{\Sigma}$}
  \Text(200,80)[b]{$\phi^{c}_{M}, \psi^{c}_{H}, \psi_{\Sigma}$}

  \Line(180,20)(220,20)    \Vertex(200,20){3}
  \Line(180,60)(220,60)    \Vertex(200,60){3}
%  \Text(270,90)[b]{$\psi^{c}_{M}, \phi^{c}_{H}, \phi_{\Sigma}$}
  \Text(270,80)[b]{$\psi^{c}_{M}, \phi^{c}_{H}, \phi_{\Sigma}$}
  \Line(250,40)(290,40)    \Vertex(270,40){3}

  \Text(35,7)[r]{$h$} \Text(40,7)[l]{- - - $\ $- - - - } \Vertex(60,7){3}
%\Line(40,7)(80,7)
  \Text(105,10)[r]{$\tilde{t}_1$} \Text(110,10)[l]{- - - $\ $- - - - }
 \Vertex(130,10){3}
%\Line(110,10)(150,10)
  \Text(105,30)[r]{$\tilde{t}_2$} \Text(110,30)[l]{- - - $\ $- - - - }
\Vertex(130,30){3}
%\Line(110,30)(150,30)
  \Text(175,10)[r]{$\tilde{t}_1^c$}  \Text(180,10)[l]{- - - $\ $- - - - }
\Vertex(200,10){3}
%\Line(180,10)(220,10)
  \Text(175,30)[r]{$\tilde{t}_2^c$}  \Text(180,30)[l]{- - - $\ $- - - - }
 \Vertex(200,30){3}
%\Line(180,30)(220,30)
\end{picture}

\caption{Tree-level KK mass spectrum of the matter, Higgs and gauge
multiplets.  Physical light Higgs and top squarks mass eigenstates are
shown in dashed lines. }
\label{Fig_spectrum}
\end{center}
\end{figure}
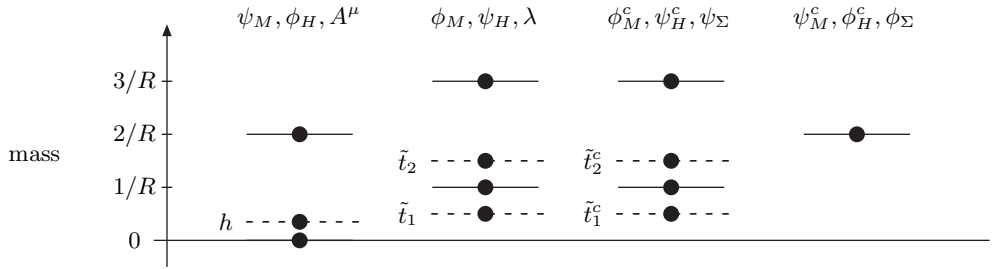

The 5D spacetime is assumed to be compactified on an $S_1/(Z_2 \times Z_2')$
orbifold.  Thus, the physical space is a line segment with
two ends -- the orbifold fixed points.  At each fixed point the 5D
fields can transform as either even or odd under the $Z_2$ symmetry
associated with that fixed point.
The field content is that of massless
${\cal N}=2$, 4D hypermultiplets and
vector multiplets.  The Kaluza-Klein (KK) reduction of this theory to 4D
yields wave functions as sines and cosines of integer and half-integer
multiplets of the size of the extra dimension, $R$, with a
mass spectrum given in Fig.~\ref{Fig_spectrum} (solid lines).

An interesting feature of this model is that the Higgs effective potential
can be calculated essentially in terms of a single free parameter $R$,
which comes from the loop contribution through the top Yukawa coupling.
The only scale in the model, $1/R$,
can be determined by the minimization condition of the
Higgs effective potential, which gives $1/R\sim{370}$ GeV.
This allows us to predict the physical Higgs boson mass as well as the
superparticle and KK tower masses.  The predicted Higgs mass is
$m_h = (127 \pm 8)~{\rm GeV}$.
At the first excited level, there are two superparticles for each SM
particle.  Their masses shift from $1/R$ by the electroweak breaking
effect through the expectation value of the Higgs field.
The largest effect appears in the top squark sector, giving top squarks
of masses $1/R \pm m_t\sim{210, 540}$ GeV.
The mass spectrum of the light Higgs and stops is
also shown in Fig.~\ref{Fig_spectrum} as dashed lines.
Although possible
additional brane localized interactions could shift
the value of $1/R$ by as much as a few tens of percent, the Higgs and stop mass
prediction is much less sensitive to the UV physics.  The observation of the
light stop, described below,
and the measurement of its mass to be in the predicted range would
be compelling evidence for this model.

\subsection{Phenomenology}

The low energy effective theory \emph{below} $\sim{2}/R$ consists of
one Higgs doublet and two superpartners for each SM particle.
The light Higgs has SM-like Yukawa couplings and $WWh$,
$ZZh$ gauge couplings.  It can be produced at Tevatron RUN II via the usual
associated production of $Wh$ or $Zh$, with the Higgs decays into $b\bar{b}$
or $\tau\bar\tau$.  At LHC, $gg\rightarrow{h}\rightarrow\gamma\gamma$
would be the discovery channel for the light Higgs.
One particularly
characteristic feature of this model is that the two degenerate
light stops ($\tilde{t}_1$), with mass
$m_{\tilde{t}_1}=1/R-m_t$,  are the LSPs,
which are stable\footnote{The cosmological constraints on an absolutely
stable stop could be relaxed if we allow a small degree of
$R$-parity violating.}
if $R$-parity is exact.
The lowest mass states \cite{oleg} are $\tplus=\tilde{t}_1\bar{d}$,
$\tzero=\tilde{t}_1\bar{u}$ and their charge conjugate states
$\tminus$, $\tzerobar$.
The other supersymmetric particles (besides the heavier stop) 
are almost degenerate at a mass of $\sim{1}/R$,
with small mass splittings coming from electroweak corrections or
additional unknown brane kinetic terms.
All of them cascade decay into stop LSPs.  

Squarks mostly decay via the channel
$\tilde{q}\rightarrow{q}\tilde{g}\rightarrow{q}\tilde{t}_1{t}
\rightarrow{q}\tilde{t}_1{b}W^{\pm}.$
The $b$-jet from the top decay and the leptons/jets from the $W$ decay
are energetic, while $q$ from the original $\tilde{q}$ decay is usually 
soft because of relatively small mass splitting between the squark and gluino. 
For $\tilde{q}_L$, another decay channel 
$\tilde{q}_L\rightarrow{q}' \tilde{W}^{\pm}
\rightarrow{q}' \tilde{t}_1{b}$
could be important (although suppressed by weak coupling) if the heavy gluino
is off-shell in the first process.

For sleptons, decay can occur via the neutral
Wino or Bino:
$\tilde{l}\rightarrow{l}\tilde{W}^0/\tilde{B}^0\rightarrow{l}\tilde{t}_1{t}
\rightarrow{l}\tilde{t}_1{b}W^{\pm}.$
However, for $\tilde{l}_L$,  decay similar to that of $\tilde{q}_L$
is comparable to the neutral gaugino process for
$m_{\tilde{W}}<m_{\tilde{l}}$, and dominates if
$m_{\tilde{W}}>m_{\tilde{l}}$.

The gluino can decay via
$\tilde{g}\rightarrow{t}\tilde{t}_1,\ b\tilde{b}_L,\ q\tilde{q}_L,\
q\tilde{q}_R$,
either on-shell or off-shell, with subsequent decays of the
$t$, $\tilde{b}_L$, $\tilde{q}_L$ or $\tilde{q}_R$ as described above.
The generic decay products would be
$\tilde{t}_1$+$b$+($l\nu/jj'$)+soft jet/lepton.
%Decays through $b\tilde{b}_L$ could lead to
%$\tilde{t}_1$+($l\nu/jj'$)+soft $b$ if the $\tilde{b}_L$
%decays via (\ref{eq:leftsbdecay}), or $\tilde{t}_1$+$b$+soft jet is also
%possible if $\tilde{q}_L$ decays via (\ref{eq:leftsqaurkdecay}).

There are three neutralinos and three charginos (all six {\it Dirac}), in contrast with
the usual MSSM with four Majorana neutralinos and two Dirac charginos.
The mass eigenstates
are usually a mixture of gauginos and Higgsinos, with a small mass splitting
of roughly 12 GeV (18 GeV) between the lightest one and the two heavier
neutralinos (charginos).  The dominant decay channel for charginos is
$\chi^{\pm}\rightarrow{b}\tilde{t}_1$,
leading to a clean signal of (track or missing energy)+$b$ jet.
Neutralinos decay similarly to gluinos or via
$\chi^0\rightarrow\chi^{\pm}W^{\mp}$ with subsequent decays of the $\chi^{\pm}$
and $W^{\mp}$.

The heavy Higgs, with mass $m_H\sim{2}/R$,
decays through $t\bar{t}$ like a usual heavy SM
Higgs, while $H\rightarrow{WW}$ is forbidden due to the
non-conservation of the fifth dimensional momentum.
The discovery of the Kaluza-Klein tower of
SM particles (heavy quarks, leptons, gauge bosons and their mirror partners)
would be strong evidence for
the existence of TeV-scale extra-dimensions.
However, they are  unlikely to be pair-produced at a TeV-scale
$e^+e^-$ collider.
At a hadron collider the cascade decay chain is complicated
and hard
to distinguish from background. Thus, the phenomenology of these heavier
states is not discussed here.

\subsection{Experimental Prospects}

Both neutral and charged $T$ mesons are stable inside the detector
since the $\beta$-decay~of $\tplus$ into $\tzero$ is suppressed by
small mass splitting between them.
$\tpm$ appears as a stiff charged track with little hadron calorimeter
activity, and passes through the muon chamber.
A heavy $\tpm$ can be distinguished from a muon
via large $d{E}/d{x}$ or time-of-flight (TOF).
$\tzero$ and $\tzerobar$ produced via cascade decays of higher-mass
SUSY states
can be identified via their missing energy
as they traverse the detector with little interaction ($\tzero$ and $\tzerobar$
from direct pair production will be difficult to trigger on).

The CDF Run I search~\cite{run1} for these types of particles
was based on the measurement of this energy loss by the tracking
system. The search found no evidence for the production of 
heavy stable charged particles (`CHAMPs'),
and lower mass limits of $\sim$200 GeV were set
in the context of a heavy quark model~\cite{hqu}. The limit in the
context of supersymmetric models with a long-lived stop is currently
being evaluated.

The CDF and D0 detectors have been upgraded in preparation for Run II.
The CDF detector includes a TOF detector.
Studies have been performed using Pythia~\cite{pyth} and the full CDF detector
simulation
to produce CHAMP Monte Carlo samples.
Simulations of the upgrade high-$p_T$ two-track trigger
indicate a 37\% efficiency for pair production of 200 GeV CHAMPs.
Combining curvature and TOF information, the
mass of 200 GeV CHAMPs can be measured with an event-by-event
resolution of about 25 GeV. In the case of D0, $d{E}/d{x}$ will be
used at the trigger level, while the muon system will provide
TOF information.

Given the 0.6 pb $\tilde{t}_1\tilde{t}_1$ cross section~\cite{tevst}
of this model, CDF would expect over 100 events
per year.
The measurement of the light stop mass in the predicted range,
with a rate consistent with twice the standard SUSY cross section,
would be compelling evidence for this model. In addition, the scalar
nature of the top squark could be inferred from its measured decay
angular distribution.

The cross section for the
production of $\tilde{q}\tilde{q}$, $\tilde{q}\tilde{g}$,
$\tilde{g}\tilde{g}$ range from $0.1-1$ pb in this model \cite{tevsquark}.
The final decay products always include two $\tilde{t}_1$, which can be
triggered on as discussed above. In addition, there will be other
activity that can be used to distinguish these events from direct
stop pair production. These generally include two energetic $b$ jets,
possible energetic leptons, missing $E_T$ (from neutrinos),
and jets coming from decays of the intermediate $W$. Alternative
triggers based on this accompanying activity
should be useful in selecting
events containing two neutral $T^0$, although distinguishing the signal
from backgrounds in this case will require further study.
%is about $0.1-1$ pb \cite{tevsquark}.  The final decay
%products always include two $\tilde{t}_1$, which can be used as a trigger as
%discussed above.  Note that in most $R$-parity conserving MSSM models with
%neutralino LSP, the
%characteristic feature of SUSY signal is $\missinget$.  While in this model,
%two stable stops (inside the detector) in the final decay products would
%always lead to 0,1,2 high $p_T$ tracks w/o $\missinget$.
%In addition, there are two energetic $b$ jets (except for the case of
%$\tilde{b}_L$),
%and possible energetic
%lepton, $\slash\hspace{-0.1 in}{E}_T$ (from $\nu$) and jets coming from the
%decay of intermediate $W$.
Finally, the observation of sleptons, gauginos and
higher mass states are less promising for Run II of the Tevatron.

Given the factor of seven increase in center of mass energy over
Tevatron Run II, the LHC experiments will substantially extend
the search for phenomena predicted by this model.  In particular, the
heavier mass states should be accessible.  
Even for the heavier stop states, hundreds of events per year can
be observed, given the clean signal of $\tilde{t}_2$ 
decay, and allowing confirmation of the robust prediction
$m_{\tilde{t}_2} - m_{\tilde{t}_1} = 2m_t$.
If the light Higgs is not
discovered at the Tevatron it will be uncovered at the LHC in the
usual channels.  The heavy
Higgs discovery should also be possible, as in the SM case, provided enough
integrated luminosity is achieved~\cite{lhc-higgs}.
However, prospects for identifying either of these Higgs as
unique to this model requires more study.  One strategy would search
for the heavy Higgs decaying via some of the predicted SUSY states.

A high-energy electron-positron Linear Collider (LC), running at
center-of-mass energies of up to 1 TeV, would permit a number
of measurements which would be essential to the confirmation
of this model. With a mass predicted to be less than 250 GeV,
pair production of the light stop would be copious even at
a 500 GeV LC. The factor-of-two enhancement of the stop production
cross section (due to the presence of the degenerate mirror state
${\tilde t}_c$) could be measured to a statistical and systematic
precision of order 1\%. At higher energies, individual thresholds for the full
spectrum of particles at the scale of $1/R$ could be detected. The
resulting picture -- a light stop plus a nearly-degenerate array of sfermions,
Higgsino states,
and gauginos, with each sfermion exhibiting a precise factor-of-two production
rate enhancement -- would represent an unambiguous signature for this model.
Another unique characteristic of this model -- the Dirac nature
of the neutralinos and charginos -- could be confirmed by studying the
helicity dependence of their production cross-sections.
The associated production of $\tilde{t}_1\tilde{t}_2$ is possible at a TeV
LC via
$s$-channel $Z$-exchange.

Several other checks of the model, which are not possible at a hadron machine,
could be performed at a Linear Collider. The $O(10\%)$
deviation from the SM top Yukawa coupling could be measured to 1-2$\sigma$
at a 1 TeV LC~\cite{tthlc}.
The chiral composition of the stop mass eigenstates could
also be confirmed by exploiting the beam polarization. All in all,
the ability to individually detect and precisely measure the properties
of each of the 1/R-scale states,
including those (such as sleptons) which are
not accessible to hadron machines, makes the complementary information provided
by the LC an essential component of the confirmation of the model's
characteristic phenomenology.

\section{The Strong CP Problem}

The vacuum of non-perturbative QCD seems to naturally support
CP violation at a level clearly at odds with basic observation.
The most well-known attempt to remove CP violation from the
QCD vacuum is known as the Peccei-Quinn mechanism~\cite{axion}, and leads
to the prediction of a massive goldstone-like boson known as
the axion. While this particle has yet to be observed,
experimental searches are only now reaching the
sensitivity needed to rule them out in mass and coupling ranges
allowed by astrophysical constraints.

Nonetheless, in recent years substantial thought has been put forward
in an attempt to explain the absence of strong CP violation
without recourse to the Peccei-Quinn mechanism. Several such models
are presented below.

\subsection{Extra Dimensions and the Strong CP Problem}

Over the past few years,
the possibility that large extra spacetime dimensions might exist
has received considerable attention. 
The main attraction of this idea is the observation
that large extra dimensions have the potential to lower the
fundamental energy scales of physics, such as the
Planck scale~\cite{ADD}, the GUT scale~\cite{DDG}, and 
the string scale~\cite{stringscale}.
In this framework, one assumes that 
the observed four dimensions are merely a subspace 
within a $p$-dimensional membrane (or D-brane) which in turn
floats in a higher-dimensional space.
The extra compactified dimensions can therefore be of two types,
either within the D-brane or transverse to it,
and the phenomenology of both types of extra dimensions  
has been explored quite extensively in recent years.

One of the surprising aspects of extra dimensions is that
they may provide a new approach towards solving the strong
CP problem.  One of the standard approaches 
to the strong CP problem is to introduce a Peccei-Quinn (PQ) symmetry
with a corresponding axion.  Unfortunately, 
the experimentally allowed parameter space for the mass and couplings of 
the axion have become very narrow, and it is not clear
how to generate the high energy scale associated with
the breaking of PQ symmetry or to explain
the ``invisibility'' of the resulting QCD axion. 

Within the framework afforded by large extra dimensions,
however, this situation may be radically altered.
The basic idea is that since the QCD axion is singlet 
under all Standard Model symmetries, the axion is free to
leave the D-brane to which all Standard-Model particles are restricted
and propagate into the higher-dimensional bulk. 
In other words, in theories with extra dimensions,
the QCD axion can accrue an infinite tower
of Kaluza-Klein axion states. 

Can this be used to lower
the fundamental PQ symmetry-breaking scale?
This issue has been investigated in Refs.~\cite{CTY,DDGaxions}.
As explicitly shown in Ref.~\cite{DDGaxions} (and first proposed in Ref.~\cite{ADD}),
it is possible to exploit the large volume factor associated with
the extra dimensions
in order to realize a large effective four-dimensional PQ scale
from a smaller, higher-dimensional fundamental PQ scale.
This is therefore one method of generating an {\it apparent}\/
high PQ symmetry-breaking scale in a natural way.

However, as discussed in Ref.~\cite{DDGaxions},
the presence of the Kaluza-Klein axions can have important and
unexpected effects on axion phenomenology.
In theories with extra dimensions,
it turns out~\cite{DDGaxions} that the 
four-dimensional axion no longer 
is a mass eigenstate;  instead,
this axion mixes with the infinite tower of Kaluza-Klein
axions, with a mass mixing matrix given in Ref.~\cite{DDGaxions}.
This mixing has a number of interesting phenomenological consequences.

First, as shown in Ref.~\cite{DDGaxions}, under certain circumstances
the mass of the axion essentially {\it decouples}\/ from the PQ scale,
and instead is set by the radius of the extra spacetime dimension.
Thus, axions in the $10^{-4}$ eV mass range are consistent with
(sub-)millimeter extra dimensions.
This decoupling implies that it may be possible to adjust the
mass of the axion independently of its couplings to matter.
This is not possible in four dimensions.

Second, as discussed in Ref.~\cite{DDGaxions},
the usual four-dimensional axion should undergo {\it laboratory  oscillations} as it propagates
since it is no longer a mass eigenstate.
Such oscillations are entirely analogous to neutrino oscillations.  
Moreover, because the axion is now
a higher-dimensional field, Standard-Model particles couple not only
to the usual four-dimensional axion zero mode, but rather to 
the entire linear superposition
$a'\sim \sum_n a_n$ (where $a_n$ are the axion Kaluza-Klein modes).
Therefore, the quantity of phenomenological interest
is the probability $P_{a'\to a'}(t)$ that $a'$ is preserved as a function of time.
It turns out~\cite{DDG} that this probability
drops rapidly from $1$ (at the initial time $t=0$) to extremely small
values (expected to be  $\approx 10^{-16}$ when an appropriately truncated
set of $10^{16}$ Kaluza-Klein
states are included in $a'$).  At no future time does this probability
regenerate.  Thus, we see that in higher dimensions,
the axion state $a'$ rapidly ``decoheres''
and becomes invisible with respect to subsequent laboratory interactions.
This decoherence is therefore a possible higher-dimensional mechanism contributing to an
invisible axion.

Finally, one can investigate the effects of Kaluza-Klein axions
on cosmological relic axion oscillations.  In this regard it is important
to understand whether the infinite towers of Kaluza-Klein axion states accelerate or retard
the dissipation of the cosmological energy density associated with
these relic oscillations.  Remarkably, one finds~\cite{DDGaxions}
that the net effect of these coupled Kaluza-Klein axions is always 
either to {\it preserve}\/ or to {\it enhance}\/ the rate of energy dissipation.
This implies that the usual relic oscillation bounds are loosened
in higher dimensions,
which suggests that it may be possible to raise the
effective PQ symmetry-breaking scale beyond its usual four-dimensional value.
This could therefore potentially
serve as another factor contributing to axion invisibility.

Together, these results suggest that
it may be possible to develop a new, higher-dimensional approach
to axion phenomenology.
Further details concerning these ideas can be found in Ref.~\cite{DDGaxions},
and other recent ideas along similar lines can be found in Ref.~\cite{otheraxions}.

\subsection{Strong CP and an Embedded MSSM}

It has been shown~\cite{babu} that a solution to the strong CP and SUSY
phase problems can be obtained if the minimal
supersymmetric model (MSSM) is embedded at high scale into a parity
conserving gauge group such as $SU(2)_L\times SU(2)_R \times U(1)_{B-L}$
or $[SU(3)]^3$. These models are motivated independently by the seesaw
mechanism for neutrino masses and due to high scale of parity restoration
maintain the nice feature of gauge coupling
unification in MSSM.

The basic idea that parity symmetry can provide a resolution of the strong
CP problem goes back to the late 1970's~\cite{goran} and is based on two
observations: (i) the first is that the tree level $\theta$ term being
parity violating
automatically vanishes in a parity invariant theory and (ii) secondly,
parity invariance makes the quark mass
matrices hermitian, thereby making that contribution to $\bar{\theta}$ i.e.
$\bar{\theta} = \theta +{\rm Arg}\{{\rm Det}(M_u M_d)\} = 0.$

The relevance of parity symmetry in solving the SUSY phase problem was
noted in \cite{rasin}, where it was shown that many of the dangerous
phases of MSSM such as the gluino mass phase, $\mu$-term phase etc all vanish due
to parity symmetry. It was then shown that loop corrections lead to a very
tiny value for $\theta$ of order $10^{-8}$ to $10^{-16}$ depending on the
structure of supersymmetry breaking terms.

\subsection{Strong CP and SUSY Non-Renormalization}

Recent developments also include
%G. Hiller and M. Schmaltz have developed
a new mechanism for solving the strong CP problem
which relies on the non-renormalization theorems of supersymmetry.
The interested reader is referred to Ref.~\cite{HS},
the content of which is discussed very briefly here.

The strong CP problem has recently become more urgent because experimental data
strongly favor a CKM phase of order one, 10 orders of magnitude larger than the
upper bound on the strong CP phase. This represents a puzzle because both
arise from the Yukawa couplings in the SM.

In this non-renormalization approach,
CP is broken spontaneously
and mediated to the SM by radiative corrections. Obtaining a large CKM phase
requires the radiative corrections to be large, i.e. strong coupling
of the CP violating physics to quarks.

Such models are very difficult if not impossible to build without
SUSY, but the non-renormalization theorem of ${\overline \theta}$ in SUSY
makes this solution to the strong CP problem very natural. In the absence of
supersymmetry breaking ${\overline \theta}$ remains exactly zero while
the CKM phase gets $O(1)$ contributions from renormalization.
If supersymmetry breaking and mediation are flavor-universal and
occur at a scale well below spontaneous CP violation
(e.g. gauge-, anomaly-, or gaugino-mediation), then the radiatively generated
${\overline \theta}$ is much smaller than the experimental bounds.
Predictions of the framework include supersymmetry with minimal flavor
violation, no CP violation beyond the large CKM phase in B physics, and nearly
degenerate first and second generation scalar masses.

Recent developments also include
an explicit model for the CP messenger sector, but
it should be stressed
that the overall framework is more general because it is based on
model-independent non-renormalization theorems.

\section{Sensitivity to Doubly Charged Higgs Bosons in the Process 
$e^-\gamma \to e^+\mu^-\mu^-$}

Doubly charged Higgs bosons would have a distinct experimental
signature. Such particles arise in many extensions of 
the Standard Model (SM) such as the 
Higgs triplet model of Gelmini and Roncadelli and the
Left-right symmetric model.  
The signals for doubly charged Higgs bosons
arising from an $SU(2)_L$ triplet were studied in the 
process $e^-\gamma\to e^+ \mu^-\mu^-$.  
Details of the analysis are given in reference \cite{us} and 
contribution P3-18 of these proceedings.
The photon was assumed to be produced by backscattering a laser from the
$e^+$ beam of an $e^+e^-$ collider \cite{backlaser}.
We consider $e^+e^-$ centre of mass 
energies of $\sqrt{s}=500$, 800, 1000, and 1500~GeV
appropriate to the TESLA/NLC/JLC high energy colliders 
and $\sqrt{s}=3$, 5, and 8~TeV for the CLIC proposal. 
In all cases an integrated luminosity of ${\cal L}=500$~fb$^{-1}$ was 
assumed. Because the signature of same sign 
muon pairs in the final state is so distinctive, with no SM background, 
the process can be sensitive to virtual $\Delta^{--}$'s with masses in
excess of the centre of mass energy, depending on the strength of the
Yukawa coupling to leptons.  

Indirect
constraints on $\Delta$ masses and couplings have been obtained from lepton
number violating processes
\cite{swartz_and_other}. Rare  decay measurements  
\cite{mu3e_and_other} yield  very stringent restrictions on the
non-diagonal couplings $h_{e\mu}$ which were consequently neglected.
Stringent limits on flavour diagonal couplings 
come from the muonium anti-muonium conversion
measurement \cite{muoniumex} which 
requires that the ratio of the Yukawa coupling, $h$, and Higgs mass, 
$M_\Delta$, satisfy
$h/M_\Delta < 0.44$~TeV$^{-1}$ at 90\% C.L..
These bounds allow the existence of low-mass doubly charged Higgs with 
a small coupling constant.  
Direct search strategies for the $\Delta^{--}$ have been explored for 
hadron colliders \cite{datta_and_others}, with the mass reach at the LHC 
extending to $\sim 850$~GeV.
Signatures have also been explored for various configurations of
lepton colliders, including $e\gamma$ colliders. 

In the process $e^-\gamma \to e^+ \mu^-\mu^-$, the signal of 
like-sign muons is distinct and SM background free, 
offering excellent potential for doubly charged Higgs discovery. 
The process proceeds via the production of a positron along 
with a $\Delta^{--}$, with the subsequent $\Delta$ decay into two muons 
as well as through additional non-resonant contributions. 
The cross section is a convolution of the backscattered 
laser photon spectrum, $f_{\gamma/e}(x)$ \cite{backlaser}, 
with the subprocess 
cross section, $\hat{\sigma}(e^- \gamma \to e^+ \mu^-\mu^-)$.
Due to contributions to the final state that proceed 
via s-channel $\Delta^{--}$'s, the doubly-charged Higgs 
boson width must be included.  The $\Delta$ width, however, 
is dependent on the parameters of the model, 
which determine the size and relative importance of various decay modes.   
To account for the possible 
variation in width without restricting ourselves to 
specific scenarios, we calculated the width using
$\Gamma (\Delta^{--}) = \Gamma_b + \Gamma_f$
where $\Gamma_b$ is the partial width to final state bosons and 
$\Gamma_f$ is the partial width into final state fermions.  
Two scenarios for the bosonic width were considered: 
a narrow width scenario with 
$\Gamma_b=1.5$~GeV and a broad width scenario with $\Gamma_b=10$~GeV. 
These choices represent a reasonable range for various values of the 
masses of the different Higgs bosons.
The partial width to final state fermions is given by
$\Gamma (\Delta^{--}\to \ell^- \ell^-) = \frac{1}{8\pi} 
h^2_{ \ell \ell} M_\Delta$.
Since we assume $h_{ ee} =h_{\mu\mu} =h_{\tau\tau} \equiv h$,
we have 
$\Gamma_f = 3 \times \Gamma (\Delta^{--}\to \ell^- \ell^-) $. Many
studies assume the $\Delta$ decay is entirely into leptons; for small
values of the Yukawa coupling and relatively low $M_{\Delta}$ this leads
to a width which is considerably narrower than our assumptions for 
the partial width into bosons. 

We consider two possibilities for the $\Delta^{--}$ signal.  We assume
that either  all three final state particles are
observed and identified or that  the
positron is not observed, having been lost down the beam pipe.  
To take into account detector acceptance we restrict 
the angles of the observed particles relative to the beam, 
$\theta_{\mu},\; \theta_{e^+}$, to the ranges $|\cos \theta| \leq 0.9$.
We restrict the particle energies 
$E_{\mu}$, $E_{e^+} \geq 10$~GeV and assumed an identification 
efficiency for each of the detected final state particles of $\epsilon = 0.9$.  

Given that the signal for doubly charged Higgs bosons is so distinctive and 
SM background free, discovery would be signaled by even one event.
Because the value of the cross section for the process we consider is
rather sensitive to the $\Delta$ width, the potential for discovery 
of the $\Delta$ is likewise sensitive to this model dependent 
parameter.  Varying $\Gamma_b$, we find that, relative 
to $\Gamma_b = 10$~GeV, the case of zero bosonic width has a sensitivity 
to the Yukawa coupling $h$ which is greater by a factor of about 5
\cite{us}.

\begin{figure}[t]
\centerline{
\begin{minipage}[t]{6.0cm}
%\begin{figure}[htbp]
\vspace*{13pt}
\centerline{
             \hspace{-0.6cm}
\includegraphics[width=5.8cm, height=6cm,angle=-90]{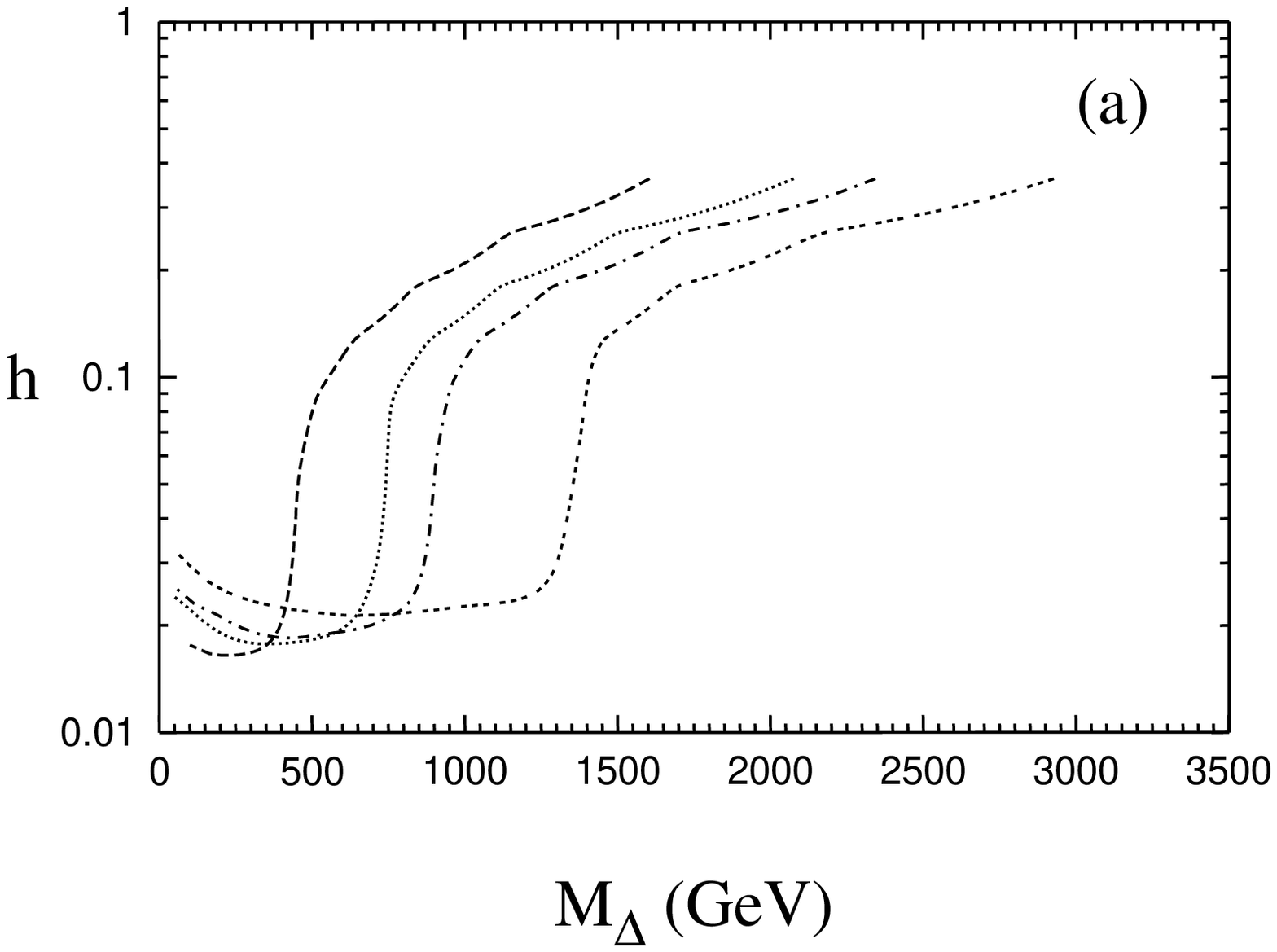}}%,width=6.1cm,clip}}
%\centerline{(6a)}
\vspace*{13pt}
\end{minipage} 
\hspace*{0.5cm}
\begin{minipage}[t]{6.0cm}
\vspace*{13pt}
\includegraphics[width=5.8cm, height=6.0cm,angle=-90]{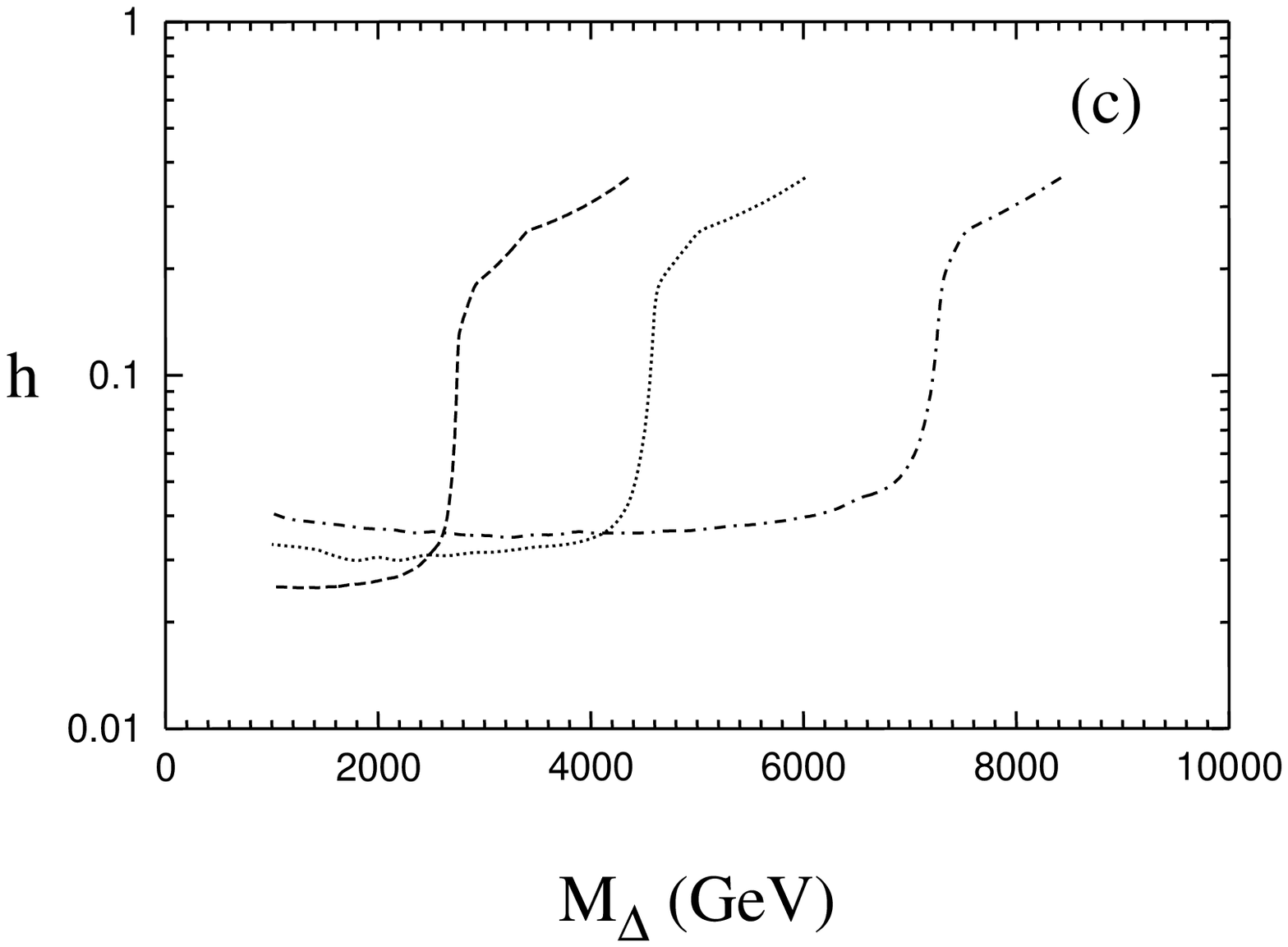}
\vspace*{13pt}
\end{minipage}
}   
\centerline{\hspace*{-0.6cm}
%\newline
\hspace*{0.5cm}
\begin{minipage}[t]{6.0cm}
\vspace*{13pt}
\includegraphics[width=5.8cm, height=6.0cm,angle=-90]{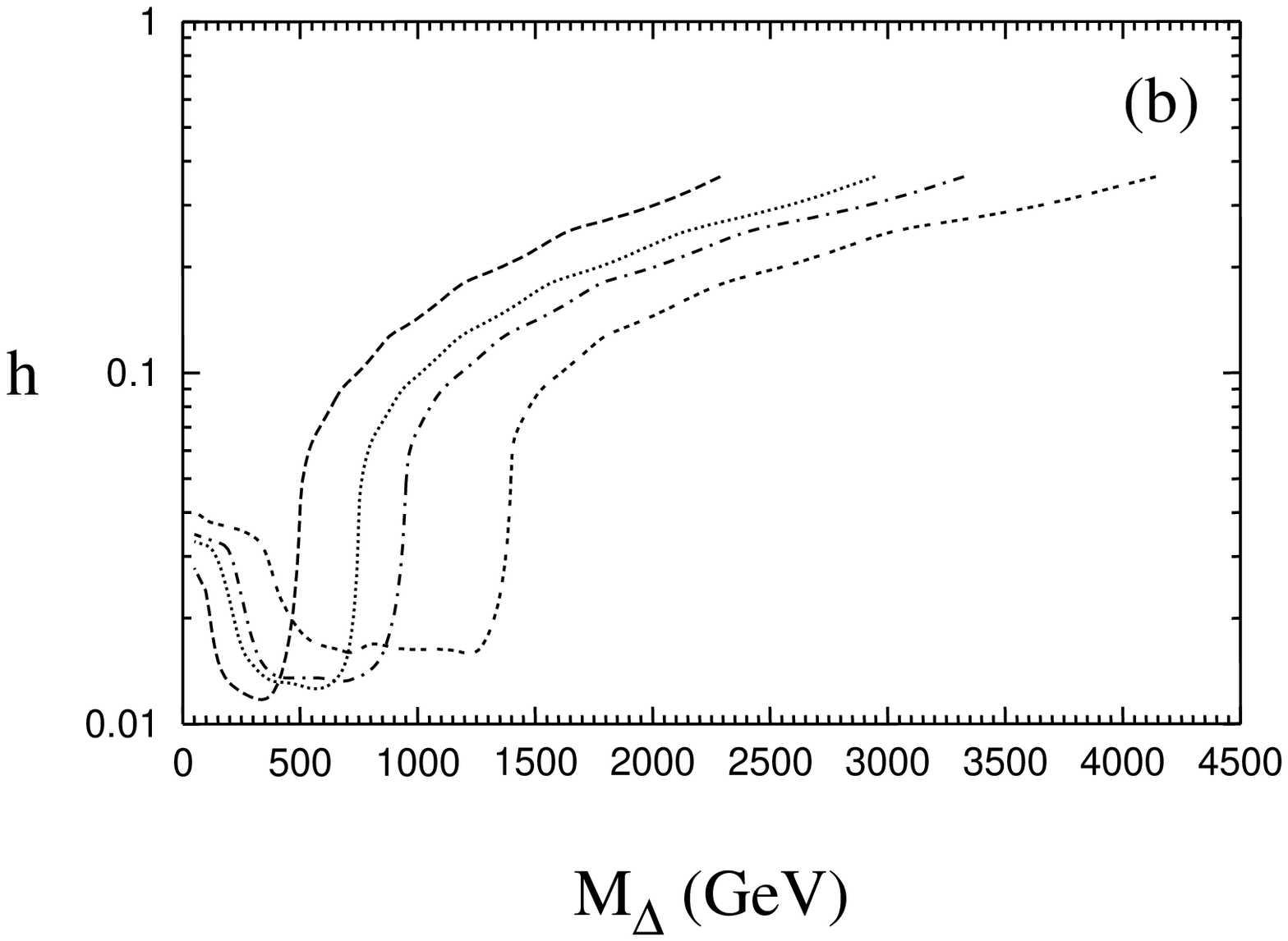}
%\centerline{(6b)}
\vspace*{13pt}
\end{minipage} 
\hspace*{0.5cm}
\begin{minipage}[t]{6.0cm}
\vspace*{13pt}
\includegraphics[width=5.8cm, height=6.0cm,angle=-90]{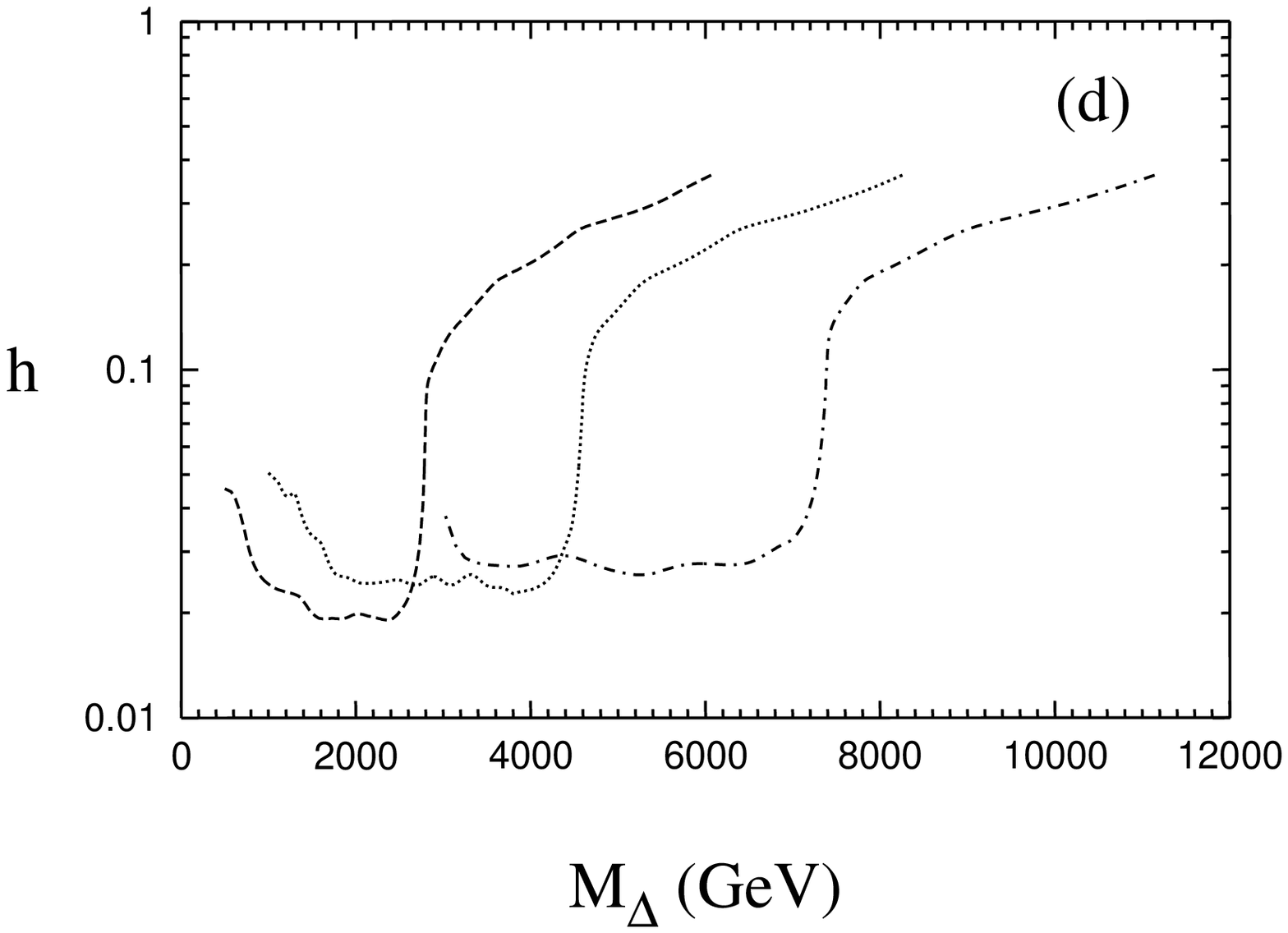}
%\centerline{(7b)}
\vspace*{13pt}
\end{minipage}
         }
\caption{  Discovery limits for the charged Higgs bosons
as a function of Yukawa coupling and $M_{\Delta}$.}
\label{dchiggs}
\end{figure}

In Fig.~\ref{dchiggs}
we show 95\% probability (3 event) contours in the $h-M_{\Delta}$ 
parameter space. In each case, we assume the narrow width 
$\Gamma=1.5+\Gamma_f$~GeV case.  
Figure \ref{dchiggs}a corresponds to the center 
of mass energies  $\sqrt{s}=500$, 
800, 1000, and 1500~GeV,  
for the case of three observed particles in the
final state, whereas Fig.~\ref{dchiggs}b 
shows the case where only the two muons are observed.
Figs.~\ref{dchiggs}c and \ref{dchiggs}d correspond to the energies being 
considered for the CLIC $e^+ e^-$ collider, namely, 
$\sqrt{s}=3$, 5, and 8~TeV,  for the three body and two body
final states, respectively.
In each case, for $\sqrt{s}$ above the $\Delta$ production threshold, 
the process is sensitive to the existence of the $\Delta^{--}$ with 
relatively small Yukawa couplings.  However, when the  $M_\Delta$ 
becomes too massive to be produced the values of the Yukawa couplings 
which would allow discovery grow larger slowly.

The observation of doubly charged Higgs bosons would represent physics
beyond the SM and, as such, searches for this type of particle
should be part of the 
experimental program of any new high energy facility.  
We found that for $\sqrt{s_{e\gamma}}> M_\Delta$ doubly 
charged Higgs bosons could be discovered for even relatively small 
values of the Yukawa couplings; $h > 0.01$. For larger values of the 
Yukawa coupling the $\Delta$ should be produced in sufficient quantity 
to study its properties.   For values of $M_\Delta$ greater than the 
production threshold, discovery is still possible due to the 
distinctive, background-free final
state in the process  $e\gamma \to e^+ \mu^-\mu^-$ which can proceed 
via virtual contributions from intermediate $\Delta$'s.  Thus, even an 
$e^+e^-$ linear collider with modest energy has the potential to 
extend $\Delta$ search limits significantly higher than can be 
achieved at the LHC.

\section{Update of Discovery Limits for Extra Neutral Gauge Bosons at 
Hadron Colliders}

Following is a discussion of discovery limits for extra neutral gauge
bosons at hadron colliders. The discovery of extra gauge bosons at future
lepton colliders, particularly those arising from string and technicolor models,
is presented in Ref.~\cite{MurWells}.

Extended gauge symmetries and the associated heavy neutral gauge 
bosons, $Z'$, are a feature of many extensions of the Standard Model
such as grand unified theories, Left-Right symmetric models, and 
superstring theories.  If a $Z'$ were discovered it would have 
important implications for what lies beyond the Standard Model.
It is therefore important to study and compare the discovery 
reach for extra gauge bosons at the various facilities 
that are under consideration for the future
\cite{cvetic,godfrey,capstick,leike,riemann,rizzo}. 
Included in the list of 
proposed facilities considered at the Snowmass'01
workshop are high energy hadron colliders. 
In this section we give the results of contribution P3-44 
\cite{p3-44} which updates previous studies 
\cite{cvetic,godfrey,capstick,rizzo}
to include the high energy hadron colliders discussed at this meeting
which range in $\sqrt{s}$ from 14~TeV to 200~TeV.

Many models that predict extra gauge bosons exist in the 
literature.  We present search limits for several of these models 
that have received recent attention. Although far 
from exhaustive, the list forms
a  representative set for the purposes of comparison.  
The Effective Rank 5 E6 Model %\cite{e6} 
starts with the GUT group $E_6$ and breaks via the chain 
$E_6 \to SO(10)\times U(1)_\psi \to SU(5)\times U(1)_\chi \times 
U(1)_\psi$ with the $Z'$ charges given by linear combinations of the 
$U(1)_\chi$ and $U(1)_\psi$ charges.
Specific models are $Z_\chi$ 
corresponding to the extra $Z'$ of $SO(10)$, $Z_\psi$ 
corresponding to the extra $Z'$ of $E_6$, 
and $Z_\eta$ 
corresponding to the extra $Z'$ arising in some superstring theories.
The Left-Right symmetric model (LRM) is based on the gauge group 
$SU(2)_L \times SU(2)_R \times U(1)_{B-L}$,
which has right-handed charged currents and restores parity at high 
energy.
The Alternative Left-Right Symmetric model (ALRM) is based on the same 
gauge group as the LRM but now arising from $E_6$ where the fermion 
assignments are different  from those of the LRM 
due to an ambiguity in how they are embedded in the {\bf 27} 
representation.
The Un-Unified model (UUM) %\cite{uum}
is based on the gauge group $SU(2)_q \times SU(2)_l 
\times U(1)_Y$ where the quarks and leptons each transform under their own 
$SU(2)$
and the KK model (KK) %\cite{kk} 
contains the Kaluza-Klein excitations of the SM gauge bosons that
are a possible consequence of theories with large extra dimensions. We 
also consider a $Z'$ with SM couplings (SSM). Details of these models 
and references are given in Ref. \cite{cvetic}.

The signal for a $Z'$ at a hadron collider consists of 
Drell-Yan production of lepton pairs with high invariant mass via 
$p p \to Z' \to l^+ l^-$.  See Ref. \cite{p3-44,capstick,godfrey,ehlq} 
for further details and references.
The cross section for $Z'$ production at hadron colliders
is inversely proportional to the $Z'$  width.  If exotic
decay modes are kinematically allowed, the $Z'$ width will become larger and
more significantly, the branching ratios to conventional fermions smaller.  
We will only consider the case that no new decay modes are allowed.

We obtain the discovery limits for this process based on 10 events 
in the $e^+e^- + \mu^+\mu^-$ channels using the EHLQ quark 
distribution functions \cite{ehlq} set 1, taking $\alpha=1/128.5$, 
$\sin^2\theta_w=0.23$, and including a 1-loop $K$-factor in the $Z'$ 
production.  Using different 
quark distribution functions results in a roughly 10\% variation in 
the $Z'$ cross sections \cite{rizzo} with the subsequent change in 
discovery limits.  
Lowering the number of events in the $e^+e^- + \mu^+\mu^-$ channels 
to 6 raises the discovery reach  about $10\%$ while lowering the
luminosity by a factor of ten  reduces the reach  by about a factor 
of 3 \cite{capstick}.
Detailed detector simulations for the Tevatron and LHC validated our 
approximations as a good estimator of the true search reach.  
Furthermore, the results of our previous studies following this 
approach are totally 
consistent with subsequent experimental limits obtained at the 
Tevatron.

In our calculations we assumed that the $Z'$ only decays into the 
three conventional fermion families.  If other decay channels were 
possible, such as to exotic fermions filling out larger fermion 
representations or supersymmetric partners, the $Z'$ width would be 
larger, lowering the discovery limits.  On the other hand, if decays 
to exotic fermions were kinematically allowed, the 
discovery of exotic fermions would be an important discovery in 
itself;  the study of the corresponding decay modes would provide additional
information on the  nature of the extended gauge structure.   

\begin{figure}
\includegraphics[width=11cm]{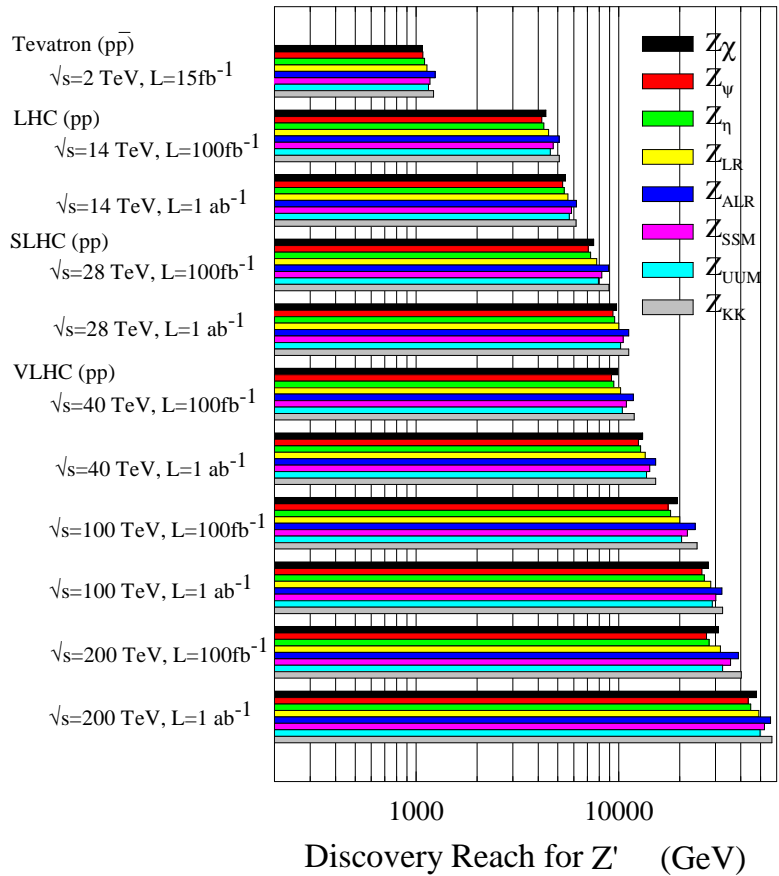} 
\caption{
Discovery limits for extra neutral gauge bosons ($Z'$) 
for the models described in the text based  
on 10 events in the $e^+e^-\ +\ \mu^+\mu^-$ channels.
}
\label{zp-pp}
\end{figure}

The discovery limits for various models at hadron colliders 
are shown in Fig.~\ref{zp-pp}.
These  bounds are relatively insensitive to specific models.  In addition, 
since they are based on a distinct signal with little background 
they are relatively robust limits.  
Typical search limits for $pp$ colliders 
are  $\sim 0.25-0.30 \times \sqrt{s}$ assuming $100$~fb$^{-1}$ to
$1$~ab$^{-1}$ of integrated luminosity with some 
variation due to differences of fermion couplings in the different 
models.
The Tevatron, a $p\bar{p}$ collider, has a 50\% higher discovery
reach than this rough guideline, 
indicating that valence quark contributions
to the Drell-Yan production process are still important at these
energies.

%\section{Summary}
\section{Parting Thoughts}

In this working group
summary we have surveyed some of the topics explored by
the Alternative Models and New Ideas subgroup.  While some of the
topics are updates of previous studies, many of the topics are
related to ideas barely imagined during the last Snowmass
workshop in 1996.  Clearly, the bounds of our creativity are challenged
only by
nature itself. Until clear experimental evidence for new
physics points us in the right direction, we must have the
capability to explore as many avenues of new physics as our
imaginations can conceive. 

The work presented hear barely scratches
the surface of the volume of alternative ideas that are currently being
pursued,
and which will no doubt continue to expand to encompass
further insights and approaches.
The challenge ahead is not only to
construct models that purport to extend our understanding of the
workings of nature, but also
to develop methods by which the predictions of these models can be tested
by experiment.  Hopefully, in the not so distant future, a
combination of theoretical creativity, phenomenological insight,
and experimental progress will peel back the next layer in our
understanding of the universe in which we live.

% figures should be put into the text as floats.
% Use the graphicx package (distributed with LaTeX2e).
% See the LaTeX Graphics Companion by Michel Goosens, Sebastian Rahtz,
% \begin{figure}
% \caption{}
% \label{}
% \end{figure}

% tables follow here or maybe be put in the text
% \begin{table}
% \caption{}
% \label{}
% \end{tabular}
% \end{table}

%\begin{acknowledgments}
%\end{acknowledgments}


\end{document}